\documentclass[aps,pra,preprint,groupedaddress]{revtex4}

\usepackage{graphicx}
\usepackage{epsfig,psfrag}
\usepackage{amsmath,amsfonts,amssymb,sistyle}

\def\be{\begin{equation}}
\def\ee{\end{equation}}
\def\bea{\begin{eqnarray}}
\def\eea{\end{eqnarray}}

\begin{document}

\preprint{S. Rasouli, S. Fathollazade, and M. Mohammadi, IASBS}
\title{Pure-amplitude holograms for high-efficiency generation of phase radial grating based radial carpet beams: Theory and experiments under plane-wave and Gaussian illumination}

\author{Saifollah Rasouli}
\email[Corresponding author: ]{rasouli@iasbs.ac.ir}
\affiliation{Department of Physics, Institute for Advanced Studies
in Basic Sciences (IASBS), Zanjan 45137-66731, Iran}
\affiliation{Optics Research Center, Institute for
Advanced Studies in Basic Sciences (IASBS), Zanjan 45137-66731,
Iran}

\author{Somaye Fathollazade}
\affiliation{Department of Physics, Institute for Advanced Studies
in Basic Sciences (IASBS), Zanjan 45137-66731, Iran}

\author{Mohammad Mohammadi}
\affiliation{Department of Physics, Institute for Advanced Studies
in Basic Sciences (IASBS), Zanjan 45137-66731, Iran}

\date{\today}

\begin{abstract}
This study introduces a pure-amplitude hologram (PAH) for generating radial carpet beams (RCBs), which are conventionally produced using pure phase radial gratings (PRGs). The hologram is designed by embedding the transmission function of a binary PRG into the phase argument of the cosine term(s) of an amplitude linear grating. When illuminated with a plane wave, this hologram generates RCBs in the non-zero diffraction orders, and when illuminated with a Gaussian beam, it generates RCB-like patterns at specific propagation distances. This method entirely eliminates the need for complex and expensive spatial light modulators (SLMs). The study presents a theory of diffraction for plane and Gaussian beams from such holograms, including a specific theoretical treatment of Gaussian beam diffraction from PRGs. Through theoretical analysis and experiments, we demonstrate how different RCBs can be generated at different diffraction orders due to the phase-amplitude enhancement resulting from multiplying the diffraction order number by the phase amplitude of the embedded base PRG, when the illuminating beam is a plane wave. For the Gaussian beam case, we show how different RCB-like patterns can be generated at different diffraction orders for the same reason, though only at specific propagation distances. Experimental and numerical results indicate that this technique yields RCBs and RCB-like patterns with approximately five times the useful power of their SLM-generated counterparts, demonstrating significantly higher power efficiency. This advantage renders the proposed method highly suitable for applications such as multiple optical trapping and free-space optical communication.
\end{abstract}

\pacs{42.00.00} Optics

\maketitle{ }

\section{Introduction}
Since their introduction in 2018, radial carpet beams (RCBs) have emerged as a distinctive class of structured light fields, prompting extensive research into their unique propagation dynamics and functional capabilities \cite{rasouli2017talbot, rasouli2018radial, hebri2018combined}. Beyond their well-known accelerating, nondiffracting, and self-healing characteristics, RCBs also exhibit local self-amplification \cite{rasouli2024power}, making them highly resilient in complex environments such as turbulent atmospheres \cite{rasouli2022resilience,rasouli2024robustness}, underwater channels \cite{Karahroudi2023Underwater}, and obstacle-perturbed paths \cite{Fathollazadeh2025SelfHealing}. These properties have enabled a range of emerging applications, including optical information transfer \cite{karahroudi2019information, Karahroudi2024Encoding} and multi-particle trapping \cite{bayat2020gear}. Parallel theoretical and experimental efforts have introduced several important extensions to the original RCB concept—such as fractional \cite{Rasouli2025fRCB}, elliptical \cite{Gao2025Elliptical,gao2026effect}, and lattice-type variants \cite{rasouli2018radial, Gong2022Petal, Masouleh2025SkewStep}—broadening their versatility and potential use cases. At the same time, innovative generation strategies based on amplitude and phase radial gratings, including sinusoidal \cite{rasouli2017talbot, Liu2024Sinusoidal}, binary \cite{rasouli2017talbot, rasouli2018radial}, synthetic-hologram \cite{Mellado2022SyntheticHolograms}, pure-phase holograms \cite{rasouli2019azimuthally}, extended Durnin’s setup \cite{Gong2025RCBlattice}, and stationary-phase designs \cite{Gong2022Petal}, have demonstrated the ability to produce nondiffracting, Bessel-like, and structurally complex RCB families such as galactic-form spinning beams \cite{rasouli2025galactic}. Together, these developments underscore the growing demand for efficient and experimentally accessible generation schemes, motivating the present work on high-efficiency phase-grating RCB generation using pure-amplitude holograms (PAHs).
%
%

An intriguing phenomenon arises when binary phase gratings are employed: upon illumination with a plane wave, high-intensity spots—whose number is twice the number of grating spokes—appear not only at the beam’s center but also across several successive concentric rings.
This behavior occurs when the phase modulation depth of the PRG reaches $\pi$ radians peak-to-peak, corresponding to a phase modulation amplitude of $\pi/2$.

However, fabricating a phase-only radial grating (PRG) with a full $\pi$-radian phase swing is technically challenging and costly. For this reason, spatial light modulators (SLMs) are often used to impose the required radial phase distribution onto a plane-wave front. Yet, due to the inherently pixelated two-dimensional (2D) lattice of SLMs, approximately 1\% of the incident power is inevitably transferred to diffraction orders even without applying any radial phase pattern \cite{moghadam2023three}. Consequently, the diffraction efficiency of RCBs generated directly from pure PRGs remains below 1\%.

Here, by adding the transmission function of a binary PRG to the argument of a pure-amplitude linear grating with a sinusoidal (or binary) transmission profile, we design and fabricate a PAH capable of generating RCBs associated with PRGs. Using the proposed method, approximately 4.38\% of the incident optical power is diffracted into the first order of the amplitude hologram, representing a significant enhancement in diffraction efficiency.
This work develops a comprehensive theory of diffraction for both plane and Gaussian beams incident upon these holograms, including a detailed analysis of Gaussian beam diffraction from PRGs. Under plane-wave illumination, the hologram transforms the incident light into RCBs within its nonzero diffraction orders. If a Gaussian beam is used instead, the hologram produces RCB-like patterns (RCBLPs), which are not propagation-invariant. For this reason, we avoid using the term RCB for them.
Although the inner part of the diffraction pattern remains invariant over a limited propagation distances, the outer region gradually loses its structure as the beam propagates. With increasing propagation distance, the region affected by structural distortion expands inward, reducing the area of invariance. Consequently, despite the appearance of stability at short ranges, the diffraction pattern evolves continuously, with changes always originating at the outer radial boundary of the patterned region.
This behavior depends on the width of the incident Gaussian beam. A broader beam enables the generated RCBLP to preserve its inner RCB-like structure over a longer propagation distance. To quantify this stability, we employ a similarity metric.
Two additional key features of this method are noteworthy. First, whether illuminated by a plane wave or a Gaussian beam, the designed amplitude hologram generates distinct RCBs or RCBLPs in all nonzero diffraction orders. This results from the phase-amplitude enhancement, where the diffraction order number multiplies the phase amplitude of the embedded base PRG, thereby amplifying the pattern. Second, the zero-order beam retains a Gaussian profile when the hologram is illuminated by a Gaussian beam. In the plane-wave case, however, edge diffraction from the hologram aperture introduces fringes into the transmitted zero-order wave.
In contrast, when a Gaussian beam directly illuminates an amplitude radial grating without the proposed phase enhancement, only the zero diffraction order is present, and an RCB-like pattern is observed only over a limited propagation range. Due to the finite width of the input beam, this pattern gradually reverts to a Gaussian profile in the far field, ultimately losing its radial structure entirely \cite{azizkhani2023gaussian}.

The ability to generate RCBs with high diffraction efficiency offers significant advantages for applications such as optical trapping and free-space optical communication.

\section{Theory and simulations}
Here we present a theoretical framework for the diffraction of plane/Gaussian beams from PAHs and its application in generating RCBs/RCBLPs via PRGs.
\subsection{Amplitude hologram diffraction: Generating RCBs/RCBLPs with embedded PRGs}
Inspired by the generation of Laguerre--Gaussian beams through the diffraction of a Gaussian beam by amplitude-only forked linear gratings~\cite{amiri2024talbot}, an amplitude-only linear--radial hologram can be constructed by implementing the transmission function of a binary-phase-only radial grating into the argument of the sinusoidally/binary modulated amplitude linear grating. The binary PRG transmission function is defined as follows~\cite{rasouli2018radial}:
\begin{equation}\label{eq:t1}
	t_{1}(\theta')=\exp\!\left(i\gamma\,\operatorname{sign}(\cos(m\theta'))\right)
\end{equation}
where, $\mathrm{sign}(\cdot)$ denotes the sign function, $m$ represents the number of spokes in the binary PRG, and $\gamma$ is the phase modulation amplitude of the phase grating.
The transmission function of a sinusoidally or binary modulated linear amplitude grating can generally be expressed in Fourier series form. In Cartesian coordinates, $(x',y')$, or equivalently in polar coordinates, $(r',\theta')$, it is written as
\begin{equation} \label{eq:t2_0}
t_{2}(x',y')=\sum_{n=-\infty}^{+\infty} t_n
\exp\!\left[-in\!\left(\frac{2\pi x'}{d}
\right)\right] ~~~ \, \text{or}  ~~~\,
t_{2}(r',\theta')=\sum_{n=-\infty}^{+\infty} t_n
\exp\!\left[-in\!\left(\frac{2\pi r'\cos\theta'}{d}\right)\right],
\end{equation}
where $t_n$ represents the Fourier coefficients, $n$ denotes the diffraction order, and $d$ is the period of the linear amplitude grating. For a sinusoidal grating, only the coefficients for $n=0$ and $n=\pm 1$ are non-zero.
%
By applying the phase-only transmission function given in Eq.~(\ref{eq:t1}) to all Fourier components of the sinusoidally or binary modulated linear amplitude grating presented in Eq.~(\ref{eq:t2_0}), the transmission function of the corresponding sinusoidal or binary amplitude-only hologram for generating RCBs/RCBLPs is obtained as follows:
\begin{equation} \label{eq:t2}
t(r',\theta')=\sum_{n=-\infty}^{+\infty} t_n
\exp\!\left[-in\!\left(\frac{2\pi r'\cos\theta'}{d}
+\gamma\,\operatorname{sign}(\cos(m\theta'))\right)\right].
\end{equation}
In the following, we show that---as Eq.~(\ref{eq:t2}) indicates---under plane-wave illumination, the diffraction patterns for the first orders (\(n = \pm 1\)) of the base grating correspond to the diffraction from the PRG described in Eq.~(\ref{eq:t1}). For higher diffraction orders, the effective phase amplitude is multiplied by the order number \(n\), resulting in patterns that correspond to different, scaled versions of the embedded PRG. This occurs because the phase profile of the PRG is replicated within every diffraction order and scales linearly with the order index.

Let us now consider the diffraction from this hologram in detail and in the general case, where the incident beam is either a plane wave or a Gaussian beam with complex amplitude

\[
U'(r',z = 0) = e^{-r'^2 / w_0^2}.
\]

Immediately after the hologram, in the polar coordinate system \((r',\theta')\) at the plane \(z = 0^+\), the optical field can be expressed as follows:

\begin{equation} \label{eq:t3}
	U'(r',\theta',0^{+})=e^{-\frac{r'^2}{w_0^2}}
	\sum_{n=-\infty}^{+\infty} t_n
	\exp\!\left[-in\!\left(\frac{2\pi r'\cos\theta'}{d}
	+\gamma\,\operatorname{sign}(\cos(m\theta'))\right)\right].
\end{equation}
For the case of a plane wave, the exponential term preceding the summation vanishes. This is mathematically equivalent to taking the limit of an infinite beam waist ($w_0 \to \infty$).

The optical field after a propagation distance of $z$ from the hologram, by using the Fresnel–Kirchhoff diffraction integral \cite{saleh2019fundamentals}, can be written as follows:	
\begin{equation}\label{eq:t4}
	U(r,\theta,z)=h\int_0^{\infty}\int_0^{2\pi}
	U'(r',\theta',0^{+})
	\exp(i\alpha(r'^2-2 r'r\cos(\theta'-\theta)))
	r'\,dr'\,d\theta',
\end{equation}
where
\begin{equation}\label{eq:t5}
	h=\frac{\exp(ikz)}{i\lambda z}e^{i\alpha r^2}, \qquad \alpha=\pi/(\lambda z),
\end{equation}
in which $\lambda$ is the wavelength and $k=2\pi/\lambda$ is the wave number. By replacing Eq.~(\ref{eq:t3}) in Eq.~(\ref{eq:t4}), the diffracted beam can be written as a summation of different diffraction orders, each propagates in a certain direction:
\begin{eqnarray}\label{eq:t6}
	&U(r,\theta,z)=
	h \sum_{n=-\infty}^{+\infty} t_n
	\int_0^{\infty}\!\!\int_0^{2\pi}e^{-r'^2(\frac{1}{w_0^2}-i\alpha)}
	\exp\!\left[-in\gamma\,\operatorname{sign}(\cos(m\theta'))\right]\\\nonumber
	&\quad\times  \exp\!\left[-in\frac{2\pi}{d}r'\cos(\theta')-2i\alpha r r'(\cos(\theta)\cos(\theta')-\sin(\theta)\sin(\theta'))\right]
	r'\,dr'\,d\theta'.
	\nonumber
\end{eqnarray}
Using $\frac{1}{w^2}=\frac{1}{w_0^2}-i\alpha$ we have
\begin{eqnarray}\label{eq:t7}
	&U(r,\theta,z)=
	h \sum_{n=-\infty}^{+\infty} t_n
	\int_0^{\infty}\!\!\int_0^{2\pi}e^{ \frac{-r'^2}{w^2}}
	\exp\!\left[-in\gamma\,\operatorname{sign}(\cos(m\theta'))\right]\\\nonumber
	&\quad \exp(-2i \alpha r' \cos \theta'\left[r \cos \theta + \frac{n\pi}{d \alpha}\right]-2i\alpha rr'\sin (\theta')\sin (\theta))
	r'\,dr'\,d\theta'.
	\nonumber
\end{eqnarray}

Now by considering parameter exchanges
$(r \cos \theta + \frac{n\pi}{d \alpha} ) = r_{n}\cos\theta_{n}$
and
$ r  \sin\theta = r_{n}\sin\theta_{n}$,
the changed coordinates for the $n$th diffraction order can be defined in the following forms:
\[
r_n=\sqrt{r^2+\left(\frac{n\pi}{d\alpha}\right)^2
	+\frac{2n\pi r\cos\theta}{d\alpha}},
\qquad
\tan\theta_n=\frac{r\sin\theta}{r\cos\theta+\frac{n\pi}{d\alpha}}.
\]
These transformations represent the shifts of different diffraction orders relative to the zero-order diffraction in the observation plane. Consequently, the complex amplitude of the beam field simplifies to:
\begin{eqnarray}\label{eq:t8}
	&U(r,\theta,z)=
	h \sum_{n=-\infty}^{+\infty} t_n
	\int_0^{\infty} e^{ \frac{-r'^2}{w^2}}r'\,dr'\,\!\!\int_0^{2\pi}
	\exp\!\left[-in\gamma\,\operatorname{sign}(\cos(m\theta'))\right]\\\nonumber
	&\quad \exp(-2i \alpha r' r_n \cos(\theta'-\theta_n)).
	d\theta'.
	\nonumber
\end{eqnarray}
By separating the different diffraction orders, this integral can be written as:
\begin{eqnarray}\label{eq:t9}
	&U(r,\theta,z)=
	h t_0 \int_0^{\infty}e^{ \frac{-r'^2}{w^2}}r'\,dr'\!\!\int_0^{2\pi} \exp\!\left[-2i\alpha r r'\cos(\theta'-\theta)\right]d\theta'+
	\nonumber \\
	&\quad h \sum_{n=1}^{+\infty} t_n \int_0^{\infty}e^{ \frac{-r'^2}{w^2}}r'\,dr'\!\!\int_0^{2\pi}e^{-in\gamma\,\operatorname{sign}(\cos(m\theta'))}\exp\!\left[-2i\alpha r' r_n\cos(\theta'-\theta_n)\right]d\theta' +
	\\\nonumber
	&\quad  h \sum_{n=1}^{+\infty} t_{-n} e^{\frac{-r'^2}{w^2}}r'\,dr'\!\!\int_0^{2\pi}e^{in\gamma\,\operatorname{sign}(\cos(m\theta'))}\exp\!\left[-2i\alpha r' r_{-n}\cos(\theta'-\theta_{-n})\right]d\theta'.
\end{eqnarray}
We express this relation based on the different diffraction orders in the following general form:
\begin{eqnarray}\label{eq:t10}
	U(r,\theta,z)=U_0(r,\theta,z)+\sum_{n=1}^{+\infty}(U_n(r_n,\theta_n,z)+U_{-n}(r_{-n},\theta_{-n},z)).		
\end{eqnarray}
Accordingly, we will investigate two cases: the complex field of the beam generated in the zero-order diffraction from the hologram, $U_0(r,\theta,z)$, and the complex field of the beam generated in the $n$-th diffraction order, $U_n(r,\theta,z)$.

Above implemented coordinate changes show that different diffraction orders propagates in different directions and the resulted diffraction patterns get distance from each other under propagation, in which over each diffraction order, the phase pattern of an RPG was implemented on the impinging beam.

It is worth noting that, in ref. \cite{rasouli2018radial} it was shown that the result of each of the integrals in Eq.~(\ref{eq:t9}) is an RCB when the incident beam is a plane wave.
Also in ref. \cite{azizkhani2023gaussian}, it was shown that under a Gaussian beam illumination of an amplitude radial grating with a sinusoidal profile, how the diffraction pattern changes under propagation. However for the case of a PRG illuminated with a Gaussian beam there is no previous study.

\subsubsection{Calculation of $U_0(r,\theta,z)$:}
We now calculate the undiffracted (zero-order) component of the diffraction pattern. This term corresponds to the portion of the field that propagates without undergoing a change in direction and is given by:
\begin{eqnarray}\label{eq:t11}
	U_0(r,\theta,z)=
	h t_0 \int_0^{\infty}e^{ \frac{-r'^2}{w^2}}r'\,dr'\!\!\int_0^{2\pi} \exp\!\left[-2i\alpha r r'\cos(\theta'-\theta)\right]d\theta'.
\end{eqnarray}	
To evaluate the angular integral, we use the integral representation of the first-kind Bessel function from Ref.~\cite{abramowitz1988handbook}, defined as $\frac{1}{2\pi} \int\limits_{0}^{2\pi} e^{i l \gamma } e^{- i x \cos \gamma} d \gamma = i^l J_l (x)$. Accordingly, Eq.~(\ref{eq:t11}) simplifies to:
\begin{eqnarray}\label{eq:t12}
	U_0(r,\theta,z)=
	t_0 h 2\pi \int_0^{\infty}e^{ \frac{-r'^2}{w^2}} J_0(2\alpha rr') r'\,dr'.
\end{eqnarray}
To evaluate this integral, we also employ an integral transformation from Ref.~\cite{gradshteyn2014table}, given as follows:
\[
\int_0^{\infty}x^{\nu +1} e^{-ax^2}j_{\nu}(bx)dx=\frac{b^\nu}{(2a)^{\nu+1}}e^{(-\frac{b^2}{4a})}.
\]
By comparing this relation with Eq.~(\ref{eq:t12}) and considering $C=h \pi w^2$ and $\alpha^2 w^2=\frac{1}{w_0'^2}$, the result of the diffraction integral simplifies to:
\begin{eqnarray}\label{eq:t13}
	U_0(r,\theta,z)=
	t_0 C e^{-\frac{r^2}{w_0'^2} }.
\end{eqnarray}
This relation indicates that the zero-order diffraction from the hologram is equivalent to the complex amplitude field of a Gaussian beam with a width of $w'_0$ and an amplitude of $t_0C$.

In the special case where the incident light is a plane wave, represented mathematically by taking the beam waist to infinity ($w_0 \to \infty$), the resulting zero-order field is likewise a plane wave.

\subsubsection{Calculation of $U_n(r_n,\theta_n,z)$}
As shown in Eq.~(\ref{eq:t9}), the complex amplitude of the field of the beam generated in a specific diffraction order $n$ is expressed as:
\begin{eqnarray}\label{eq:t14}
	&U_n(r_n,\theta_n,z)=\\\nonumber
	&h t_n \int_0^{\infty}e^{ \frac{-r'^2}{w^2}}r'\,dr'\!\!\int_0^{2\pi}e^{-in\gamma\,\operatorname{sign}(\cos(m\theta'))}\exp\!\left[-2i\alpha r' r_n\cos(\theta'-\theta_n)\right]d\theta'.
\end{eqnarray}
This integral describes the calculation of the diffraction of a Gaussian beam from a PHG. On the other hand, Ref.~\cite{azizkhani2023gaussian} presents the general relation for the diffraction of a Gaussian beam from radial structures as follows:
\begin{equation} \label{eq:t15}
	U(r,\theta,z)=Ce^{-R^2}(t_0+\sum_{q=1}^{+\infty}(t_q e^{iq\theta}+t_{-q}e^{-iq\theta}) \mu_q(R)).
\end{equation}
In this relation, $R$ is a dimensionless parameter defined as $R=\frac{\pi r}{\lambda z}w$. Furthermore, $t_q$ represents the Fourier expansion coefficients of the radial structure, and $C$ is equal to $\pi hw^2$. Additionally, $\mu_q(R)$ is expressed in terms of the modified Bessel functions of the first kind of order $\frac{q\pm1}{2}$, denoted by $I_{\frac{q\pm 1}{2}}(.)$, as follows:
\begin{equation} \label{eq:t16}
	\mu_q(R)=\frac{\sqrt{\pi}}{2} (-i)^q R {e^{\frac{R^2}{2}}}[I_{\frac{q-1}{2}}(\frac{R^2}{2})-I_{\frac{q+1}{2}}(\frac{R^2}{2})].
\end{equation}
On the other hand, for a PRG with a binary profile and the transmission function given in Eq.~(\ref{eq:t1}), the Fourier expansion and its coefficients have been calculated in Ref.~\cite{rasouli2018radial} as follows:
\begin{eqnarray}\label{eq:t17}
	t_1(\theta')=\cos(\gamma)+\sum_{l=1(odd)}^{\infty} \frac{2}{l\pi} i^l \sin(\gamma)(e^{iml\theta'}+e^{-iml\theta'}).
\end{eqnarray}
By comparing this relation with the general transmission function of a radial structure,
$
t_1(\theta')=\sum_{q=-\infty}^{\infty} t_q e^{iq\theta'} =t_0+\sum_{q=-\infty}^{\infty}(t_q e^{iq\theta'}+t_{-q}e^{-iq\theta'}),
$
the Fourier coefficients for a PRG are obtained as $t_0=\cos(\gamma)$ and $t_{q=ml}=t_{q=-ml}= i^l \frac{2}{l\pi} \sin(\gamma)$.
Substituting these coefficients into Eq.~(\ref{eq:t15}), the diffraction integral of the Gaussian beam from the binary PRG is derived as follows:
\begin{equation} \label{eq:t18}
	U(r,\theta,z)=Ce^{-R^2}\{\cos(\gamma)+\sum_{l=1(odd)}^{+\infty} i^l \frac{4}{l\pi} \sin(\gamma)\cos(ml\theta) \mu_{ml}(R)\}.
\end{equation}
According to Eq.~(\ref{eq:t14}), for a given diffraction order $n$, the phase amplitude is scaled by a factor of $n$, becoming $n\gamma$. By applying this to Eq.~(\ref{eq:t18}), the diffraction integral for each positive, non-zero order $n$ of a Gaussian beam incident on the amplitude-only hologram yields the following relation:
\begin{equation} \label{eq:t19}
	U_n(r_n,\theta_n,z) =  t_n C_n e^{-R_n^2}[\cos(n\gamma)+\sum_{l=1(odd)}^{+\infty} i^l \frac{4}{l\pi} \sin(n\gamma)\cos(ml\theta_n) \mu_{ml}(R_n)].
\end{equation}
In this relation,
\[
R_n = \frac{\pi w r_n}{\lambda z}
\quad \text{and} \quad
C_n = \frac{e^{ikz}e^{i\alpha r_n^2}}{1 + i\frac{z}{z_R}},
\]
where \( z_R = \frac{\pi w_0^2}{\lambda} \) is the Rayleigh length.
Accordingly, the general diffraction relation for a Gaussian beam from the proposed pure amplitude hologram becomes:
\begin{equation} \label{eq:t20}
	U(r,\theta,z) =\sum_{n=-\infty}^{+\infty} \left( C_n t_n e^{-R_n^2}[\cos(n\gamma)+\sum_{l=1(odd)}^{+\infty} i^l \frac{4}{l\pi} \sin(n\gamma)\cos(ml\theta_n) \mu_{ml}(R_n)]\right).
\end{equation}
In this relation, \( t_n \) are the Fourier coefficients of the base linear grating that constitutes the amplitude hologram, and their values depend on the grating's duty cycle.
Because it is both impractical and unnecessary to fabricate a PAH with a base linear grating having an exact duty cycle of 0.5, the Fourier coefficients of the corresponding embedded PRG must be determined experimentally. This is done using the specific PAH in question, following a method that will be discussed in detail in the Experimental section.

For both plane wave and Gaussian beam illumination, when the linear grating profile is considered sinusoidal, its Fourier series has only three non-zero terms $n=0, \pm 1$. As a result the diffraction order are three, one is zero order and two  $\pm 1$ orders.

Finally, by considering the illuminating beam as either a plane wave or a Gaussian beam, the resulting intensity pattern can be written in the form:
\[
I(r,\theta,z) = U(r,\theta,z) \, U^*(r,\theta,z),
\]
where the complex optical field \( U(r,\theta,z) \) is given by Eq.~(5) in Ref.~\cite{rasouli2018radial} for plane wave illumination, and by Eq.~\eqref{eq:t20} for Gaussian beam illumination. It is worth noting that even in the case of plane wave illumination, Eq.~\eqref{eq:t20} remains applicable by taking the limit of the beam waist to infinity (\( w_0 \to \infty \)).

\begin{figure*}[t]
	\centering
	\includegraphics[width=0.85\linewidth]{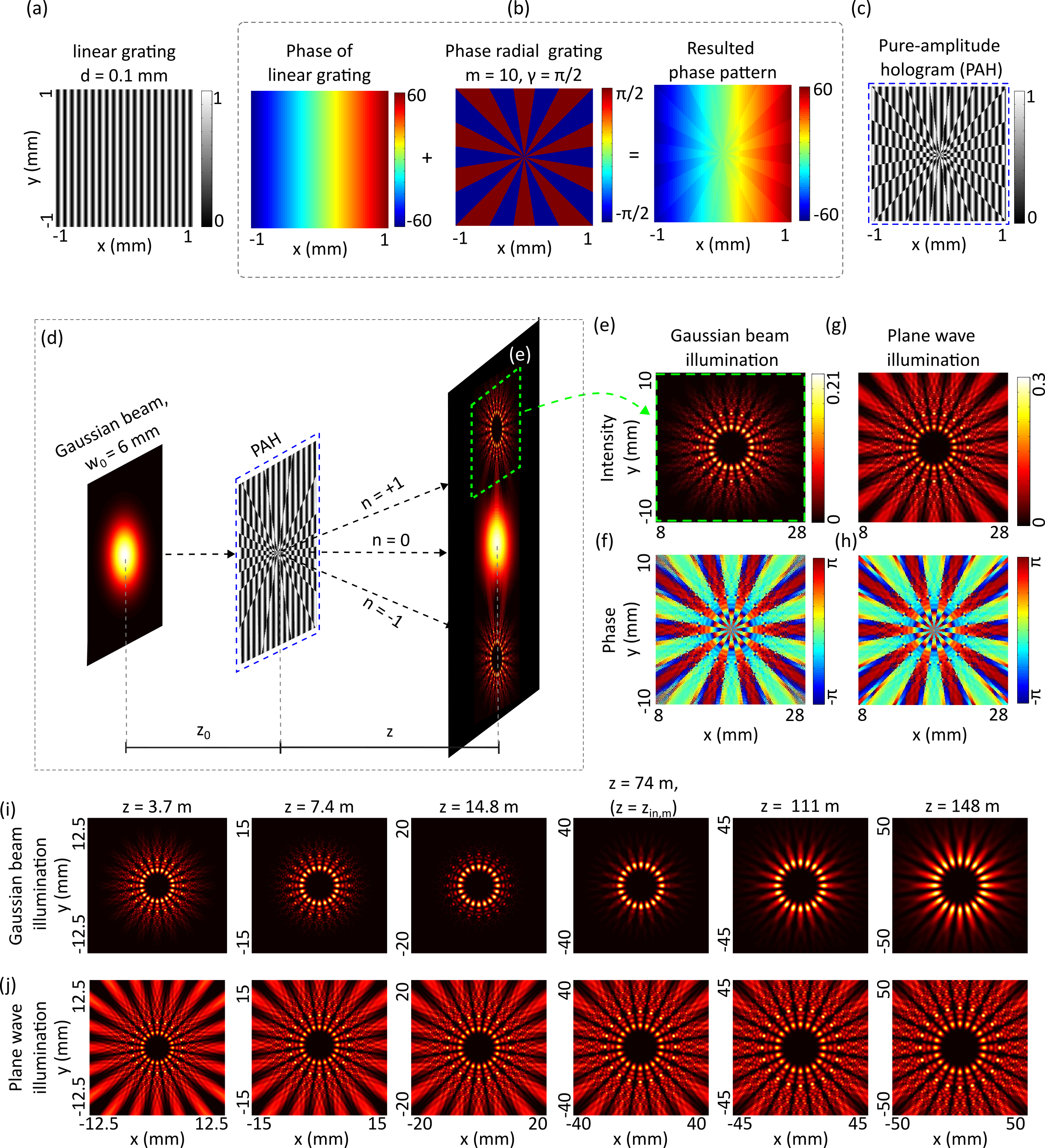}
	\caption{Design steps of a PAH and the generation of an/a RCB/RCBLP using it.
(a) Transmission function of a linear sinusoidal amplitude grating with a spatial frequency of 10~lines/mm.
(b) Addition of the phase of a PRG with $m=10$ spokes and a phase modulation depth of $\gamma=\pi/2$ to the phase function of the linear grating, forming the desired hologram phase.
(c) Amplitude hologram for generating an RCB.
(d) Illumination of the hologram with a Gaussian beam of width $w_0 = 8$~mm, producing RCBs in the $+1$ and $-1$ diffraction orders. The propagation distance between the hologram and the observation plane is $z = 2.5$~m, ensuring sufficient separation of the diffraction orders.
(e) and (f) Intensity and phase distributions, respectively, of the RCB generated in the $+1$ diffraction order.
(g) and (h) present the intensity and phase profiles, respectively, of the RCB generated in the $+1$ diffraction order when the PAH is illuminated by a plane wave.
(i) and (j) Intensity distributions of RCBLPs and RCBs generated during propagation in the $+1$ diffraction order, respectively.}
\label{fig:Fig1}
\end{figure*}

Figure~\ref{fig:Fig1} illustrates the complete design procedure for a PAH capable of generating a PRG–based RCB/RCBLP. As depicted in Fig.~\ref{fig:Fig1}(a), the design starts with a linear sinusoidal or binary amplitude grating having a spatial frequency \(d\), for example, 10~lines/mm. To incorporate the desired radial phase structure, the phase profile of a PRG with \(m = 10\) spokes and a modulation depth of \(\gamma = \pi/2\) is superimposed onto the phase function of the linear grating. This results in the composite hologram phase shown in Fig.~\ref{fig:Fig1}(b). Replacing the linear grating phase with this composite phase yields the final PAH structure presented in Fig.~\ref{fig:Fig1}(c).
When the hologram is illuminated, it generates the intended RCBs/RCBLPs in the \(+1\) and \(-1\) diffraction orders, which become spatially well separated after propagating a distance of \(z = 2.5\)~m. The specific beam produced depends on the illumination profile. Under Gaussian beam illumination with a waist radius of \(w_0 = 8\)~mm, the hologram generates RCBLPs, whose intensity and phase distributions in the \(+1\) order are shown in Figs.~\ref{fig:Fig1}(e) and (f), respectively. In contrast, under plane-wave illumination, the hologram produces ideal RCBs, with their intensity and phase profiles presented in Figs.~\ref{fig:Fig1}(g) and (h), confirming successful beam reconstruction for this illumination scenario.
A fundamental distinction between these two beam types lies in their propagation behavior, as depicted in Figs.~\ref{fig:Fig1}(i) and (j). While the RCBs generated by plane-wave illumination remain invariant under propagation (Fig.~\ref{fig:Fig1}(j)), the RCBLPs produced by Gaussian illumination exhibit a clear evolution of their structure (Fig.~\ref{fig:Fig1}(i)). This behavior can be understood from Eq.~(\ref{eq:t20}), where the coefficient \(C_n\) depends on both \(r_n^2\) and \(z\) in a distinct manner, unlike the other terms which are functions solely of the parameter \(R\). Under Gaussian illumination, the intensity distribution additionally conforms to a Gaussian envelope, matching the profile of the illuminating beam.
The structural evolution of the RCBLP can be characterized by two key radial boundaries. The outer boundary, defined by the outermost observable intensity spots, scales with the incident beam waist as \(a w_0\), where \(a = 1.4\) is a dimensionless factor determined empirically from the intensity distribution. Concurrently, the inner boundary of the pattern expands with propagation according to \(r_{\text{in}} = \sqrt{m \lambda z / \pi}\). The propagation distance at which this expanding inner region reaches the outer boundary marks a critical transition in the beam's structure. Setting \(a w_0 = r_{\text{in}}\) yields this characteristic distance:

\begin{equation}\label{eq:critical_distance}
a^2 w_0^2 = r_{\text{in}}^2 = \frac{m \lambda z_{\text{in},m}}{\pi}, \qquad \Rightarrow \qquad z_{\text{in},m} = \frac{a^2 \pi w_0^2 / \lambda}{m} = \frac{a^2 z_R}{m},
\end{equation}
where \(z_R = \pi w_0^2 / \lambda\) is the Rayleigh range.
For the experimental parameters corresponding to Fig.~\ref{fig:Fig1}(i)—\(w_0 = 8\) mm, \(\lambda = 532\) nm, and \(m = 10\)—this characteristic distance is calculated to be approximately 74 m. Beyond this propagation distance, the overall structure of the RCBLP undergoes significant alteration. It is precisely this deviation from the ideal propagation invariance of conventional RCBs that motivates designating the Gaussian-illuminated pattern as an RCBLP, highlighting its nature as a radial-carpet-like beam whose shape evolves under propagation.

All MATLAB codes used to generate the theoretically predicted patterns presented in this work are publicly available in Ref. \cite{rasouli_open-source_2026}.


\begin{figure*}[htbp]
	\centering
	\includegraphics[width=0.85\linewidth]{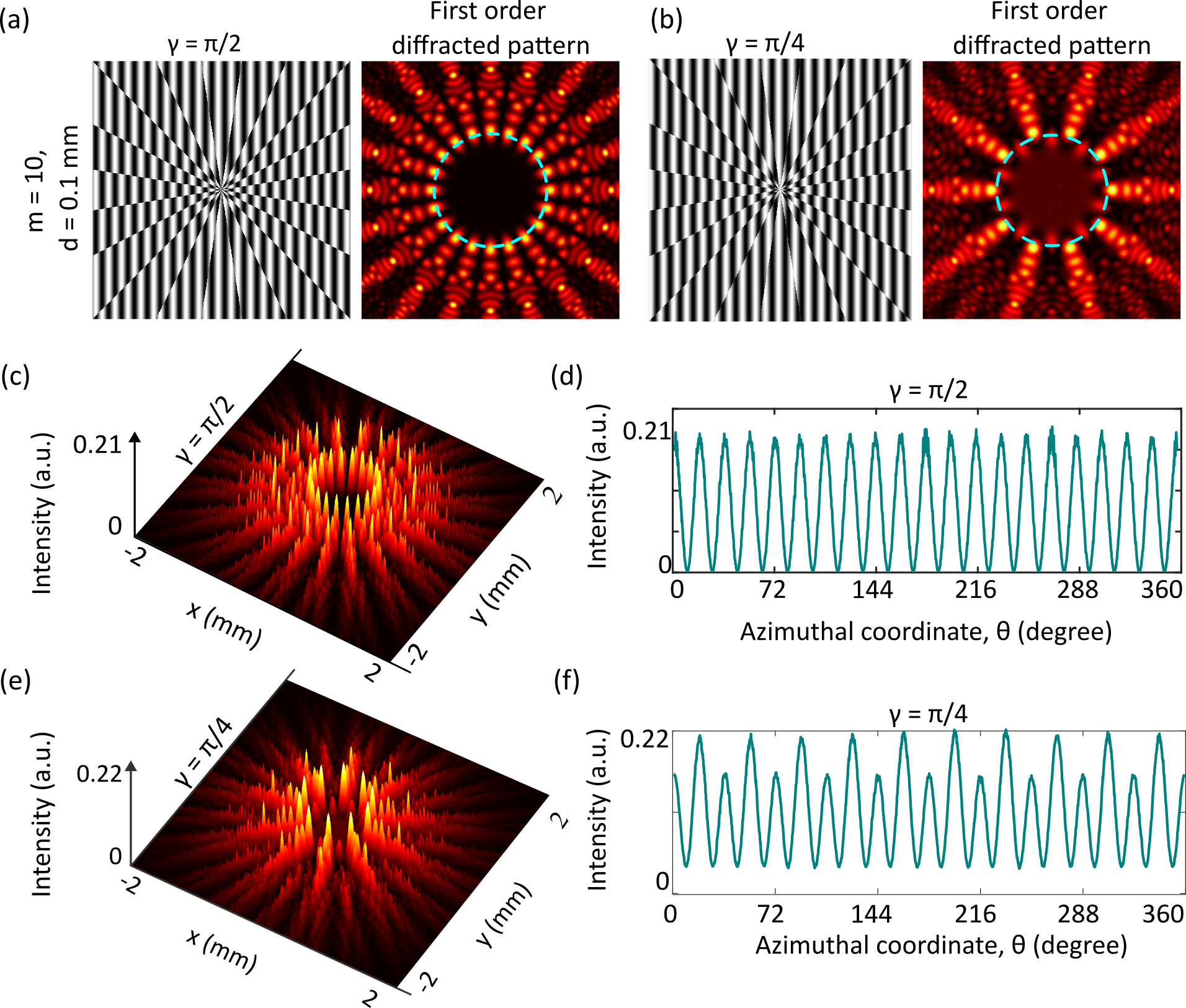}
\caption{
Investigation of the influence of the phase modulation depth $\gamma$ on the structure of the PAH and on the RCBLP generated in the first diffraction order.
(a) Transmission function of the amplitude hologram for $\gamma = \pi/2$ with $m = 10$ spokes, along with the corresponding intensity distribution of RCBLP.
(b) Same as (a), but for $\gamma = \pi/4$.
(c) and (e) Three-dimensional representations of the RCBLP produced by the holograms with $\gamma = \pi/2$ and $\gamma = \pi/4$, respectively.
(d) and (f) 1D intensity profiles extracted along circles passing through the MISs of the RCBLPs for $\gamma = \pi/2$ and $\gamma = \pi/4$. These circles are indicated by blue dashed lines in panels (a) and (b).
In both cases, the underlying linear grating has a spatial frequency of 10~lines/mm.}
\label{fig:Fig2}
\end{figure*}

Figure~\ref{fig:Fig2} examines the effect of the phase modulation depth $\gamma$ on both the structure of the amplitude hologram and the characteristics of the resulting RCBLPs in the first diffraction order. Figures~\ref{fig:Fig2}(a) and \ref{fig:Fig2}(b) show the transmission functions of PAHs designed with $\gamma = \pi/2$ and $\gamma = \pi/4$, respectively, each incorporating a PRG with $m = 10$ spokes. The corresponding RCBLP intensity distributions demonstrate how reducing the phase modulation depth modifies the contrast of the main intensity spots (MISs) of the generated beams.

To further illustrate these differences, Figs.~\ref{fig:Fig2}(c) and \ref{fig:Fig2}(e) present three-dimensional views of the RCBLP intensity patterns associated with the two modulation depths. Additionally, the one-dimensional (1D) intensity profiles shown in Figs.~\ref{fig:Fig2}(d) and \ref{fig:Fig2}(f) are extracted along circular traces passing through the MISs of the RCBLPs, highlighted by the blue dashed circles in Figs.~\ref{fig:Fig2}(a) and \ref{fig:Fig2}(b). These profiles provide a clearer comparison of the radial modulation and uniformity of the MISs for the two values of $\gamma$. In both configurations, the base linear grating maintains a spatial frequency of 10~lines/mm.

These images, along with the intensity profiles over the MISs, show that when $\gamma = \pi/2$, a total of $2m$ MISs are formed along the central ring of the beam, each with equal brightness and a maximum visibility of 1.

\subsection{Propagation Analysis and Acceptable Invariance Range for RCBLPs from a PAH}
We analyze the propagation of PAH-generated RCBLPs to determine a practical invariance criterion, showing that although the invariance is not absolute, an acceptable range can be defined within which the diffracted patterns are effectively invariant.
\subsubsection{Determination of the propagation distance \( z_1 \) for separation of diffraction orders}
Here, we calculate the minimum propagation distance required to separate different diffraction orders. For a linear grating with a period $d$, the diffraction angle of the $n$th order is given by \cite{saleh2019fundamentals}:
\begin{equation}\label{eq:t21}
d \sin \theta_n = n \lambda .
\end{equation}
For an optical grating with a sufficiently small period, the diffraction angle for the $n$th order can be estimated by
\[
\tan \theta_n \simeq \sin \theta_n \simeq \theta_n = \frac{y_n}{z},
\]
where $y_n$ represents the transverse distance of the center of the $n$-th diffraction order from the center of the zero order at a propagation distance $z$.

To determine the minimum distance at which the diffraction orders are completely separated, we consider the incident beam as a Gaussian beam and the PAH with a sinusoidal profile. As explained previously, upon illuminating the hologram with a Gaussian beam, the zero-order diffraction corresponds to the Gaussian beam and the $n=\pm 1$ orders correspond to RCBLPs. Therefore, the condition for complete separation of the orders can be expressed as follows:
\begin{equation}\label{eq:t22}
y_{\pm 1,z_1}(z) = 2w(z=z_1),
\end{equation}
where, the Gaussian beam width at the propagation distance $z_1$---the distance at which the diffraction orders are fully separated---is denoted by $w(z=z_1)$.
In the present work, for propagation distances near the hologram plane, $w(z=z_1)$ can be approximated by the beam width at the hologram plane, $w_0$.
Here, $z_1 = 2.8~\text{m}$ and the Rayleigh length is $z_R = 212.5~\text{m}$.
Since $z_1 \ll z_R$, this approximation is justified.

Consequently, the condition for the complete separation of the diffraction orders—or in other words, the distance at which the orders are completely separated—is obtained as follows:
\begin{equation}\label{eq:t23}
z_1 = \frac{2 d w_0}{\lambda}.
\end{equation}

\begin{figure}[ht!]
\centering
\centerline{\includegraphics[width=.85\linewidth]{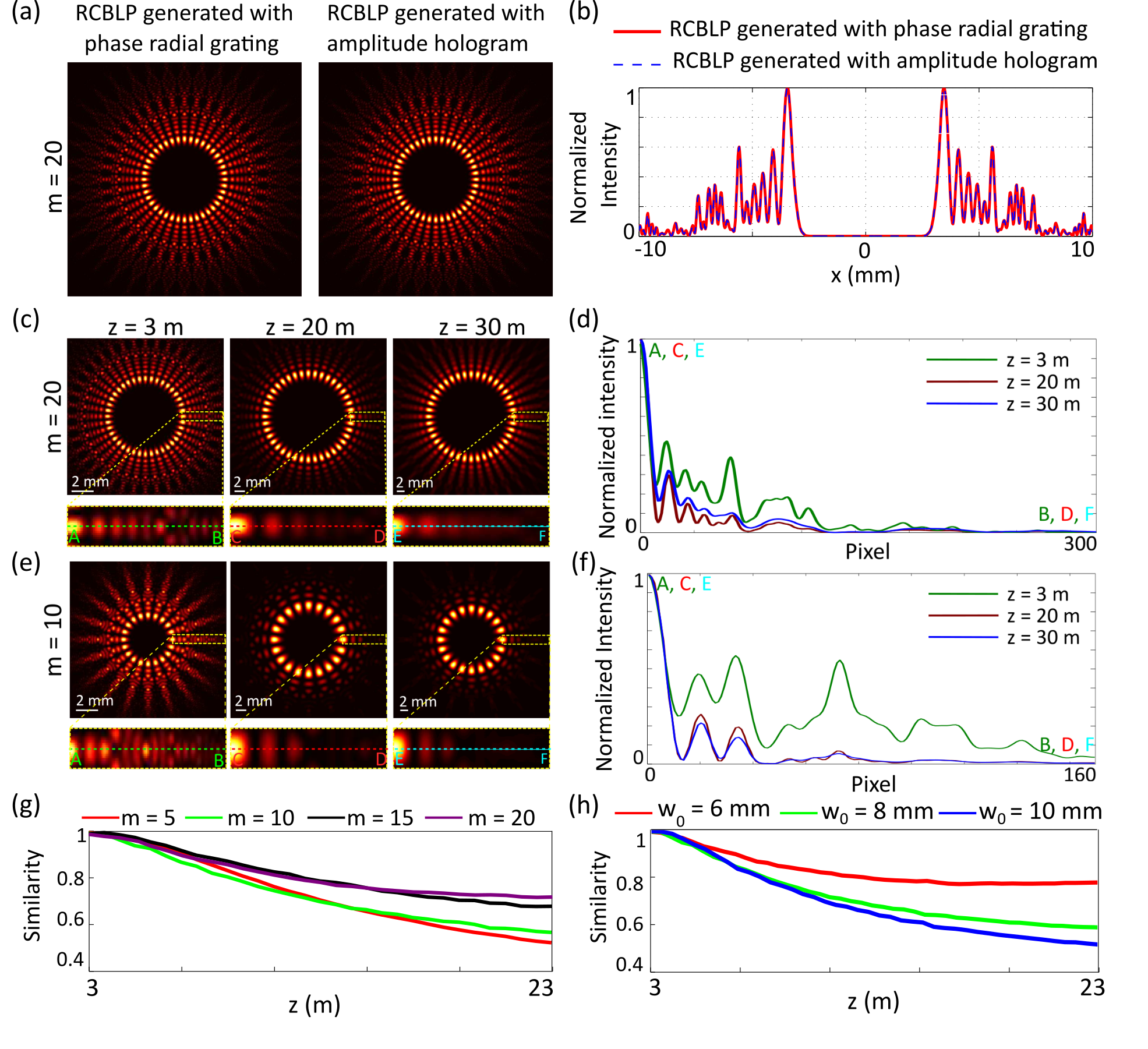}}
\caption{Investigation of the propagation and invariance of the RCB generated in the $+1$ diffraction order using a sinusoidal amplitude hologram.
(a) Comparison of the intensity distribution of the RCB generated in the $+1$ diffraction order using the amplitude hologram with that obtained directly from a binary PRG with $m=20$. The waist of the incident Gaussian beam illuminating both structures is $8~\mathrm{mm}$.
(b) 1D radial intensity profiles extracted from the beam centers in panel~(a). The two profiles show excellent agreement.
(c), (e) Intensity distributions of RCBs with $m=20$ and $m=10$, respectively, at three different propagation distances from a hologram with a base grating period of $8~\mathrm{lines/mm}$. The insets show magnified views of the regions marked by the yellow dashed boxes in panels~(c) and~(e).
(d), (f) Corresponding 1D radial intensity profiles extracted from the enlarged regions in panels~(c) and~(e), respectively, at the same three propagation distances. The radial extent of the profiles at each propagation distance is re-scaled with respect to that at the reference distance.
(g), (h) 1D similarity function as a function of the propagation distance $z$ for beams generated using the amplitude hologram, illustrating the effect of the number of spokes ($m$) and the waist of the incident Gaussian beam on the propagation-invariant length of the RCB.
}
\label{fig:Fig3}
\end{figure}

Figure.~\ref{fig:Fig3}(a) illustrates the intensity pattern cross-section in the longitudinal plane under propagation, showing the paths of different diffraction orders when a Gaussian beam with a width of $w_0 = 6$ mm is diffracted through a sinusoidal amplitude hologram with $\gamma = \pi/2$, $m = 20$, and a period of 8 lines/mm. The white dashed lines indicate three distinct propagation distances: before the separation of the diffraction orders ($z = 1.5$ m), the optimal distance at which the orders are completely separated ($z = z_1 = 2.8$ m), and $z = 4$ m, where there is free space between the order patterns. Figure~\ref{fig:Fig3}(b) shows the 2D transverse intensity distributions at these three propagation distances. As previously mentioned, the intensity patterns generated in the non-zero diffraction orders are RCBLPs. An important consideration is the propagation distance over which the RCBLPs generated in the diffraction orders remain almost shape-invariant.
In the central region of the RCBLPs, a set of $2m$ intensity spots—which we term MISs—appear with equal brightness when the condition $n\gamma = \pi/2$ is met. When this condition is not satisfied, two interleaved sets of spots with unequal intensities emerge.

There are other rings with the same number of intensity spots around the central intensity ring. After passing a certain number of such rings, the number of spots doubles on the next intensity ring \cite{masouleh2025skew}. We utilized this region to define the almost propagation-invariant feature of the diffraction patterns; that is, the pattern is considered almost shape-invariant as long as these spots having a $4m$ spot number are distinguishable in the intensity distribution. The reason for selecting this region as the almost shape-invariant region is that the spots are arranged in a regular and repeating pattern within the cylindrical coordinate system \cite{masouleh2025skew}. Figure~\ref{fig:Fig3}(c) displays the intensity distribution of an RCBLP generated in the first diffraction order, and Fig.~\ref{fig:Fig3}(d) shows a magnification of the area marked by the green dashed box in (c).

To clearly illustrate the boundary between the nearly shape-invariant central region and the diffracting peripheral region, we plot an azimuthal intensity profile at an angle of $30^\circ$, corresponding to the green and blue lines in Fig.~\ref{fig:Fig3}(e). This profile allows us to count the number of MISs in two distinct zones: the central region, where $2m$ spots of equal intensity appear (green line along A--B), and the region where the number of spots has increased to $4m$ (blue line along C--D).
Figure~\ref{fig:Fig3}(f) shows the azimuthal profiles along the C--D line at $30^\circ$ for several propagation distances. As the beam propagates, the maximum intensity of the spots in this region decreases. This attenuation results from the rapid radial expansion of the pattern and the non-shape-invariant behavior of the RCBLPs.
For comparison, under plane wave illumination, the resulting RCBs exhibit invariant intensity profiles along these spots during propagation, a key characteristic that qualifies them as true beams \cite{rasouli2018radial}.


\subsubsection{Assessing the Near-Invariance of PAH-Generated RCBLPs with a Similarity Function}

To quantify the near-invariance of PAH-generated RCBLPs, we introduce a similarity function. After generating the diffraction pattern with the PAH, we examine the propagation dynamics and near-invariant behavior of the RCBLP produced in the $+1$ diffraction order. We begin by comparing the intensity distribution of this propagated RCBLP with that of an RCBLP that was fully generated but not subjected to further propagation.

\begin{figure}[t]
	\centering
	\includegraphics[width=.85\linewidth]{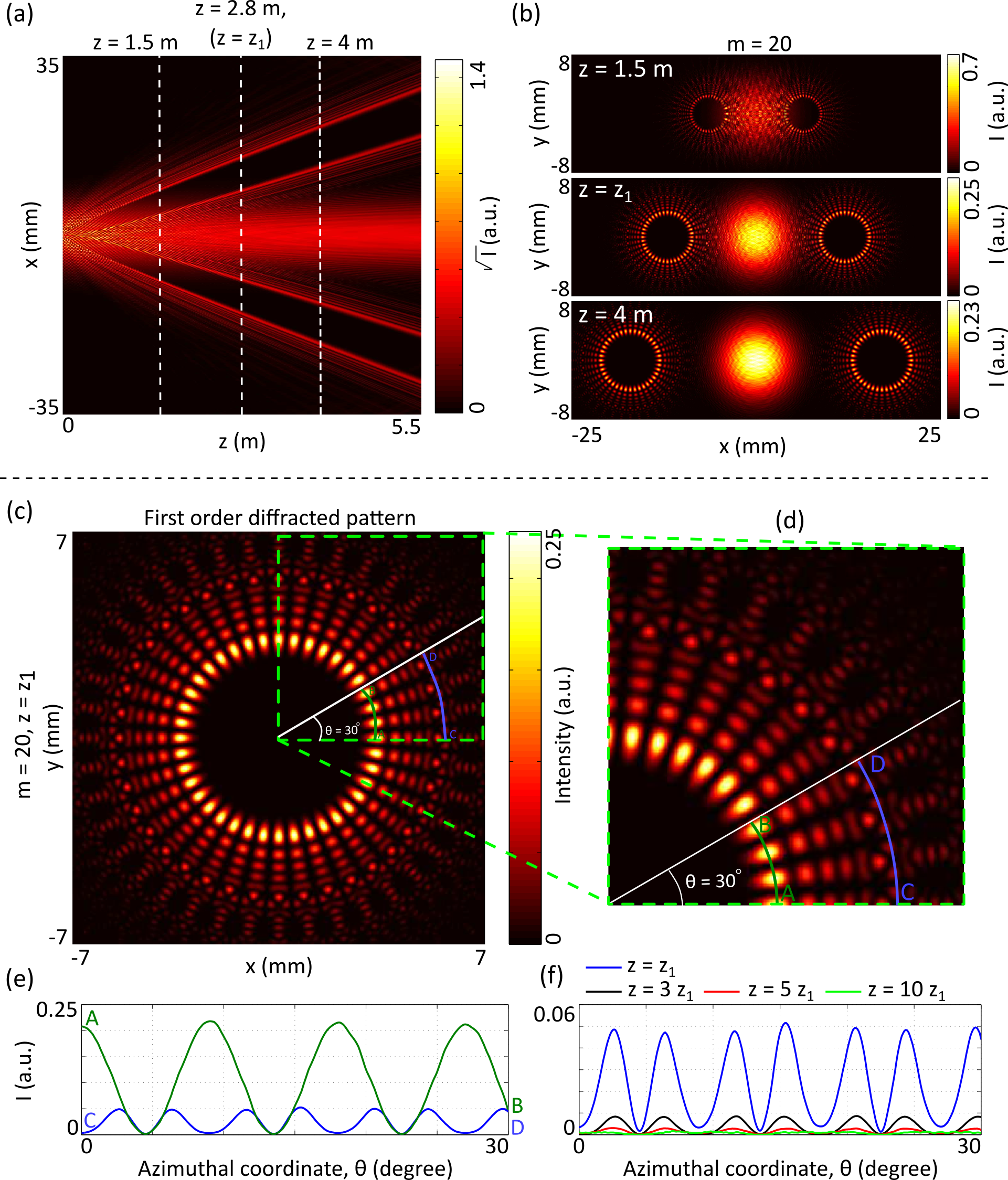}
\caption{
(a) Longitudinal cross-section of an RCBLP under Gaussian beam illumination ($w_0 = 6$~mm) after passing through a sinusoidal amplitude hologram with parameters $\gamma = \pi/2$, $m = 20$, and a spatial frequency of 8~lines/mm. The white dashed lines indicate three key propagation distances: $z = 1.5$~m (diffraction orders not yet separated), $z = z_1 = 2.8$~m (complete separation of orders), and $z = 4$~m.
(b) Transverse intensity distributions at these three propagation distances.
(c) Intensity distribution of the RCBLP in the first diffraction order at $z = z_1$.
(d) Magnified view of the region marked by the green box in part (c).
(e) Azimuthal intensity profiles over a $30^\circ$ arc along two distinct lines: the blue line A--B traversing the MISs and the green line C--D passing through a region where the number of spots has doubled.
(f) Azimuthal intensity profiles along the C--D line at various propagation distances, illustrating the evolution of the intensity distribution in this region.
}
	\label{fig:Fig4}
\end{figure}

First, we compare two RCBLPs: one obtained by diffracting a Gaussian beam (width $w_0 = 8$~mm) from the PAH, and the other from a binary PRG with $m = 20$ and $\gamma = \pi/2$. As shown in Fig.~\ref{fig:Fig4}(a), the intensity patterns of the two patterns are practically identical. Figure~\ref{fig:Fig4}(b) presents the corresponding 1D radial intensity profiles, which exhibit complete overlap.

An important consideration for diffraction from a PAH with a sinusoidal base grating is that, as noted earlier, only the $0$ and $\pm 1$ diffraction orders are generated. Initially, these orders overlap, and as the propagation distance increases, they gradually separate. It is therefore essential to determine the distance over which the beam in the $+1$ (or $-1$) diffraction order remains nearly invariant.

As mentioned above, in the intermediate region of an RCBLP, an equal number of identical intensity spots appear along each concentric ring. After a certain number of such rings, the number of intensity spots abruptly doubles~\cite{Masouleh2025SkewStep}. We use two criteria to assess the invariance of the RCBLP.

First, we employ a qualitative criterion based on the intensity patterns within the ring where this doubling occurs: as long as these intensity spots remain visible in the RCBLP's intensity distribution, the RCBLP is considered invariant. This region is chosen because the spots form a highly regular and repeating pattern in cylindrical coordinates~\cite{Masouleh2025SkewStep}.

Second, to quantitatively evaluate the RCBLP invariance, we use a similarity function computed between 1D intensity profiles, defined as
\begin{equation}\label{eq:t25}
S(z) = \frac{\sum_{i} I(r_i,z) I(r_i,z_1)}{\sum_{i} I(r_i,z)^{2} + \sum_{i} I(r_i,z_1)^{2} - \sum_{i} I(r_i,z) I(r_i,z_1)}.
\end{equation}
In Eq.~(\ref{eq:t25}), $I(r,z)$ denotes the 1D radial intensity profile of the RCBLP at propagation distance $z$, taken along the radial direction passing through the intensity maxima. This profile is compared with the corresponding profile at a reference distance, $I(r,z_1)$. The index $i$ runs over discrete radial positions. The reference distance $z_1$ is defined as the minimum propagation distance at which the diffraction orders are completely separated. This distance depends on the period of the linear grating and the waist of the illuminating beam. For example, for a hologram with a grating period of $8~\text{lines/mm}$ and an incident Gaussian beam with a waist of $w_0 = 8~\text{mm}$, the reference distance is $3~\text{m}$.

Figures~\ref{fig:Fig4}(c) and \ref{fig:Fig4}(e) show the intensity distributions of RCBLPs with $m=20$ and $m=10$, respectively, at three different propagation distances from a hologram with a grating period of $8~\text{lines/mm}$. The insets display magnified views of the regions indicated by the yellow dashed boxes in panels (c) and (e). Panels (d) and (f) present the corresponding 1D radial intensity profiles at the same three propagation distances for holograms with $m=20$ and $m=10$, respectively.

Figures~\ref{fig:Fig4}(g) and \ref{fig:Fig4}(h) compare the 1D similarity function to examine the influence of the number of spokes in the radial grating and the waist of the incident Gaussian beam on the propagation invariance of RCBLPs. As the propagation distance increases, the transverse size of the pattern expands; therefore, to enable a meaningful comparison of the intensity profiles, the propagated intensity distribution at each distance is re-scaled relative to that at the reference distance.

The similarity curves indicate that increasing the waist of the incident Gaussian beam allows the RCBLP to remain nearly invariant over longer propagation distances. Moreover, for smaller values of $m$, the near-invariance is preserved even at larger propagation distances due to the smaller size of the RCBLP's central region.

\section{Experimental results and comparison with theory and simulation}
\begin{figure}[ht!]
\centering
\centerline{\includegraphics[width=0.85\linewidth]{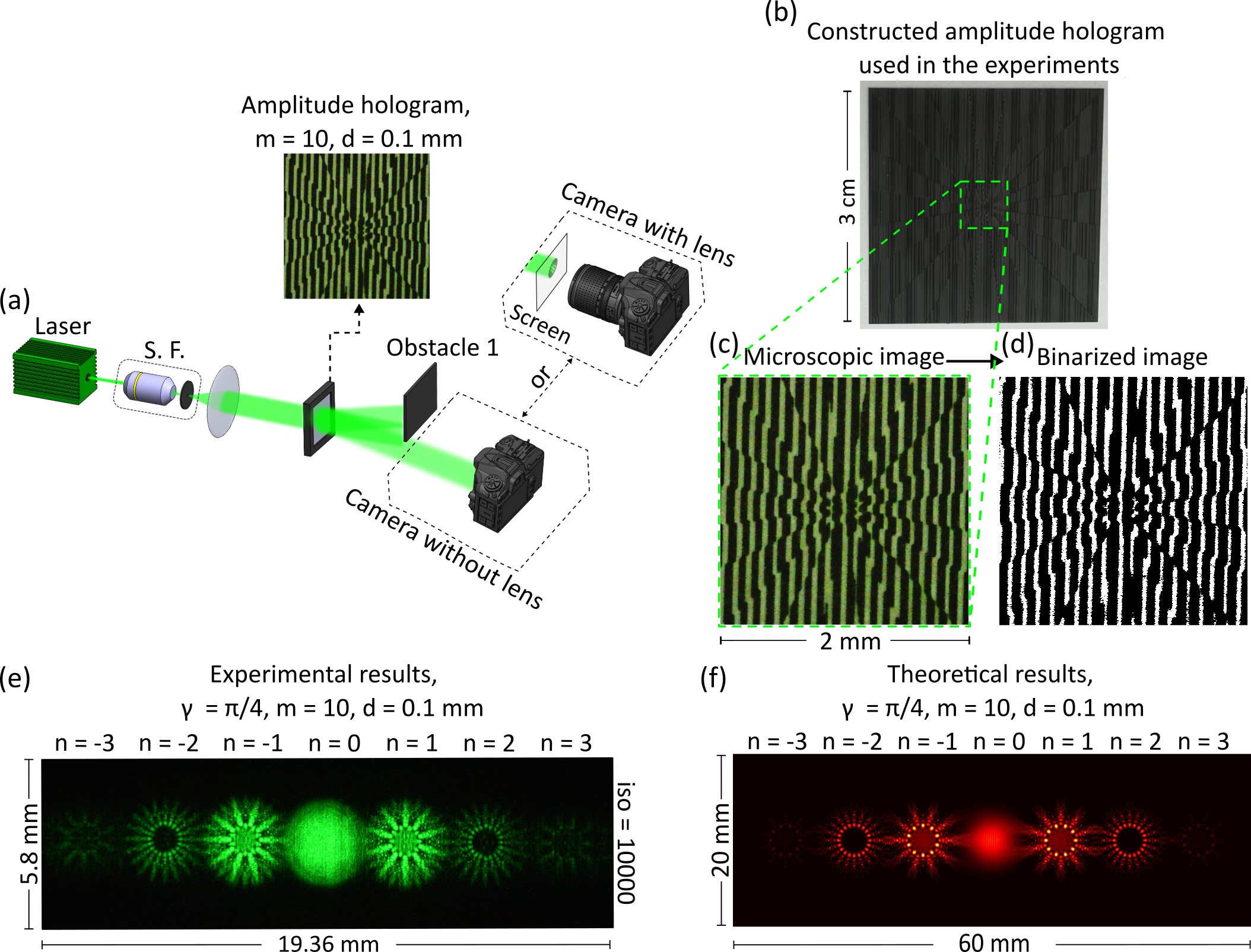}}
\caption{
(a) Experimental setup for generating RCBs using an amplitude-only hologram.
(b) Image of the amplitude hologram used, with dimensions of $3\times3$ cm, a period of $d=0.1$ mm, and a phase modulation coefficient of $\gamma=\pi/4$.
(c) Microscopic image of the magnified region of the hologram shown in part (b).
(d) Binarized image of the microscopic image.
(e) Recorded pattern of different beams generated by the amplitude hologram at a distance of 1.5 m, captured using a camera equipped with a lens.
(f) Theoretical intensity logarithm calculated at a distance of $z=1.5$ m from the hologram based on Eq.~\ref{eq:t20} for the diffraction of a Gaussian beam with a width of $w_0=3.5$ mm, using a hologram with a phase modulation coefficient of $\gamma=\pi/4$ and a period of 0.1 mm.
While the actual dimensions of the pattern in (e) on the diffuse paper screen were comparable to those in (f), the dimensions captured on the optical sensor differ after imaging with the camera lens, as illustrated in the figure.
}
\label{fig:Fig5}
\end{figure}

The experimental setup used in this work is shown schematically in Fig.~\ref{fig:Fig5}(a). A diode-pumped Nd:YAG laser beam with a wavelength of 532 nm passes through a spatial filter and is collimated using a lens to illuminate the binary amplitude hologram.
%
%
The desired beam is generated and propagates in the $+1$ diffraction order.
An image of the fabricated amplitude hologram is shown in Fig.~\ref{fig:Fig5}(b). For better clarity, a magnified image of its central region, recorded with a microscope, is presented in Fig.~\ref{fig:Fig5}(c), and its binarized version is shown in Fig.~\ref{fig:Fig5}(d).
As discussed in the theory section, achieving an exact duty cycle---such as 0.5---for the base linear grating during fabrication is neither strictly feasible nor necessary for device functionality. However, because the optical response, particularly the Fourier coefficients $t_n$ of the embedded PRG, is highly sensitive to this parameter, the actual duty cycle must be determined experimentally from the fabricated structure.
To obtain the duty cycle, we analyzed the microscopic image in Fig.~\ref{fig:Fig5}(c), which clearly reveals the binary nature of the grating. This image was binarized into two levels (0 and 1), as shown in Fig.~\ref{fig:Fig5}(d), and the duty cycle was calculated as the ratio of the area corresponding to value 1 to the total area. From this analysis, an average duty cycle of $0.27$ was obtained. The base linear grating period $d = 10~\text{mm}^{-1}$ was also measured using two complementary methods: direct measurement at multiple points on the microscopic image, and far-field measurement of the diffraction orders.

With the average duty cycle determined, the corresponding Fourier coefficients $t_n$ of the base linear grating were calculated. From these coefficients, the power fraction directed into the first diffraction order was obtained theoretically as $4.79\%$ for $\gamma = \pi/2$ and $5.05\%$ for $\gamma = \pi/4$. These theoretically derived power shares are essential for accurately modeling the diffraction behavior of the PAH.
A comparison between the experimental results and theoretical predictions based on these calculated coefficients shows good agreement. Experimentally, for $\gamma = \pi/2$, we measured the ratio of power transferred to the first diffraction order relative to the total incident power before the hologram as $4.38\%$.

Two imaging methods were used to record the intensity patterns of the generated beams. In the first, a lensless camera was placed directly in the beam path. All diffraction orders except the desired $+1$ order were blocked, allowing only its intensity distribution to be recorded. In the second method, all diffraction orders were allowed to impinge on a diffuser, and the resulting intensity distribution was imaged using a camera with its lens.

To compare the efficiency of the proposed method, beam generation using a spatial light modulator (SLM) was also performed. The power of the incident Gaussian beam before impinging on the PAH or the SLM was measured to be $10.03~\text{mW}$. In the case of the PAH, the measured power in the first diffraction order was $0.44~\text{mW}$, corresponding to approximately $4.38\%$ of the incident power. In contrast, when using the SLM, the measured power in the first diffraction order was $0.096~\text{mW}$, which corresponds to about $0.95\%$ of the input power. This reduced efficiency arises because the SLM constitutes a 2D periodic structure, causing the transmitted light to be distributed among a large number of diffraction orders. Consequently, the power of the beam generated using the amplitude hologram is approximately $4.61$ times higher than that obtained using the SLM.

The hologram was fabricated on a plastic substrate using a lithographic printing technique. Owing to the limited printing resolution in the experimental implementation, a binary hologram was used. As a result, the intensity pattern recorded by the camera, shown in Fig.~\ref{fig:Fig5}(e), contains a larger number of diffraction orders compared to the case of a sinusoidal grating.
For comparison, in Fig.~\ref{fig:Fig5}(f) corresponding theoretical patterns are also presented.
It is noteworthy that when the phase modulation depth is set to $\gamma = \pi/4$, according to Eq.~(\ref{eq:t2_0}), each diffraction order $n$ acquires an effective phase amplitude equal to $n \times \gamma$. For example, when $\gamma = \pi/4$, the effective phase amplitude in the second diffraction order becomes $2\gamma = 2\pi/4 = \pi/2$. As a result, the beam generated in the second diffraction order exhibits twice the number of central intensity spots. This behavior is clearly observable in \ref{fig:Fig5}(e) and (f).

\begin{figure*}[ht!]
	\centering
	{\includegraphics[width=0.85\linewidth]{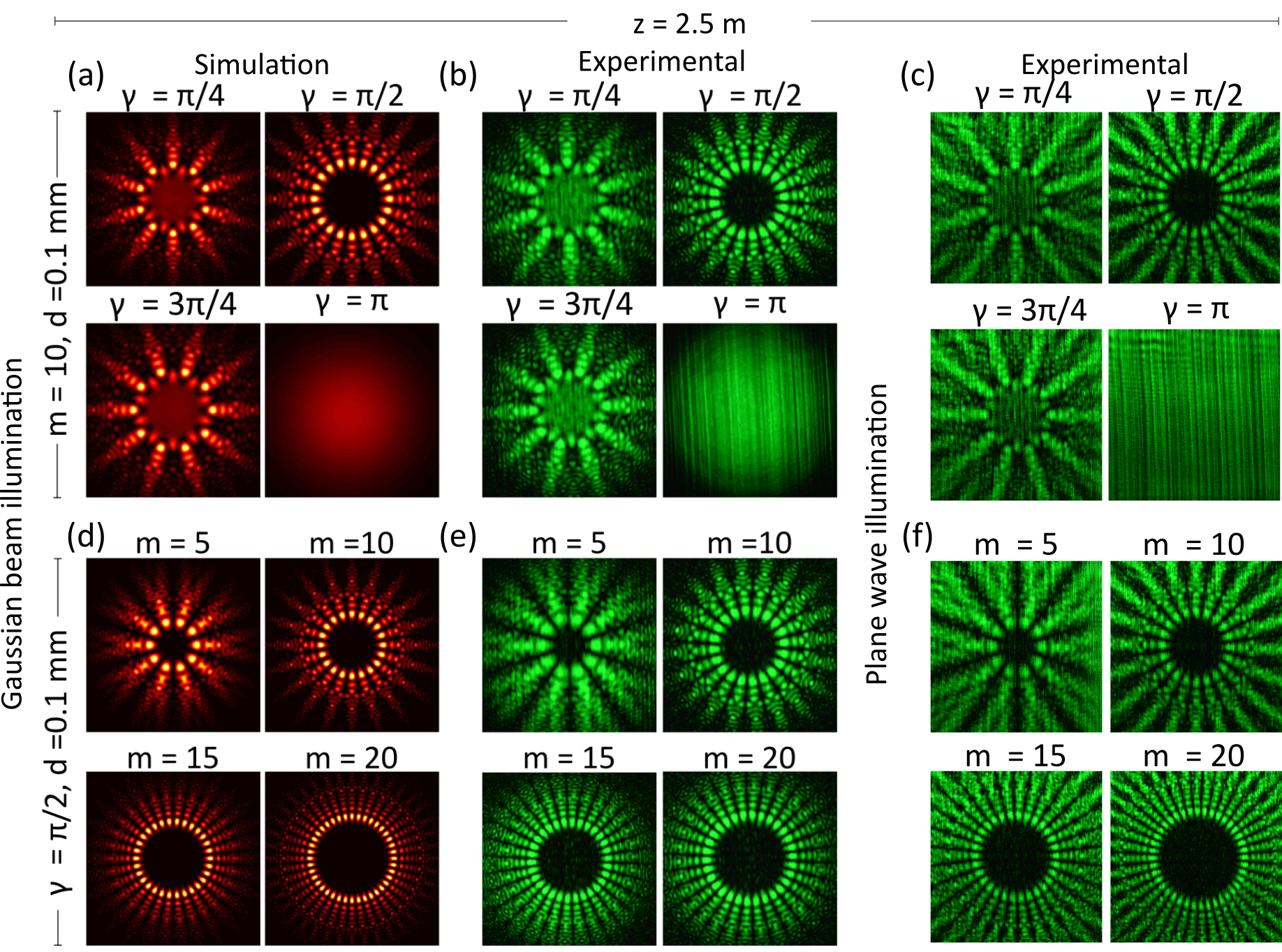}}
	\caption{Simulated and experimental intensity distributions of RCBLPs and RCBs. Parts (a) and (b) show the simulated and experimental intensity distributions, respectively, of RCBLPs generated in the first diffraction order from a PAH with a period of 0.1~mm, $m = 10$ spokes, and various modulation depths $\gamma$, recorded at a distance of $z = 2.5$~m from the hologram. Part (c) displays the experimental intensity patterns of RCBs obtained under plane-wave illumination for the same parameter set. Parts (d) and (e) depict the simulated and experimental intensity distributions, respectively, for RCBLPs produced in the first diffraction order from a PAH with a fixed modulation depth $\gamma = \pi/2$ and varying numbers of spokes $m$, also recorded at $z = 2.5$~m. Part (f) presents the corresponding experimental results for RCBs generated under plane-wave illumination.}
	\label{fig:Fig6}
\end{figure*}

Figure~\ref{fig:Fig6}(a) and (b) present the simulated and experimental intensity distributions, respectively, of RCBLPs generated in the first diffraction order from a PAH with a period of 0.1~mm, $m = 10$ spokes, and various modulation depths $\gamma$, recorded at a distance of $z = 2.5$~m from the hologram. Figure~\ref{fig:Fig6}(c) shows the experimental intensity patterns of RCBs obtained under plane-wave illumination for the same set of parameters.
Similarly, Figs.~\ref{fig:Fig6}(d) and (e) depict the simulated and experimental intensity distributions, respectively, for RCBLPs produced in the first diffraction order from a PAH with a fixed modulation depth $\gamma = \pi/2$ and varying numbers of spokes $m$, also recorded at $z = 2.5$~m. Figure~\ref{fig:Fig6}(f) displays the corresponding experimental results for RCBs generated under plane-wave illumination.
As seen, a good agreement between simulation and experiment is evident for all cases, with the plane-wave illumination results matching the simulated RCB profiles presented in Fig.~\ref{fig:Fig1}(j).

\begin{figure*}[ht!]
	\centering
	{\includegraphics[width=0.85\linewidth]{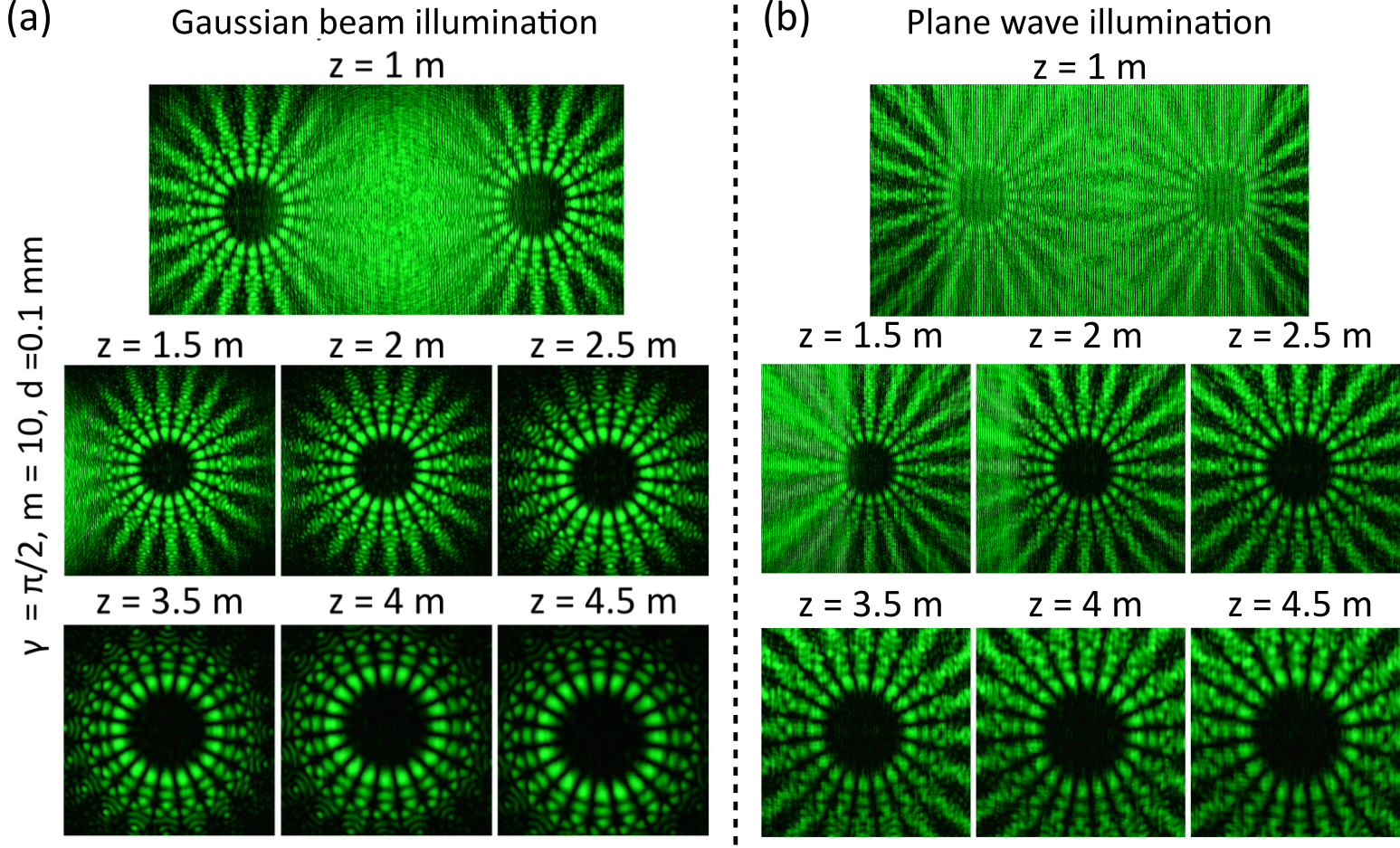}}
	\caption{Propagation of diffracted patterns at varying distances. (a) Intensity distribution of RCBLPs generated in the first diffraction order from a PAH with parameters $\gamma=\pi/2$, $m=10$, and a period of $0.1~\text{mm}$ along the propagation direction. (b) Intensity distribution of RCBs produced by the diffraction of a plane wave from the same holographic structure at different propagation distances. While the RCBs remained unchanged upon propagation, the RCBLPs exhibited slight variations, predominantly in their higher radial regions.}
	\label{fig:Fig7}
\end{figure*}

Figure~\ref{fig:Fig7}(a) illustrates the intensity distribution of RCBLPs generated in the first diffraction order from a PAH with parameters $\gamma=\pi/2$, $m=10$, and a period of $0.1~\text{mm}$ along the propagation direction. Figure~\ref{fig:Fig7}(b) presents the intensity distribution of RCBs produced by the diffraction of a plane wave from the same holographic structure at various propagation distances.
As observed, the RCBLPs exhibited marginal alterations during propagation, primarily confined to the peripheral radial region. In contrast, the RCBs remained invariant throughout propagation.
In the experimental setup, two different beam conditions were employed. For the results shown in part (a), a doublet collimating lens with a focal length of 20 cm was used to generate a Gaussian beam. For part (b), a spherical singlet collimating lens with a focal length of 1 m was used to produce a wider Gaussian beam, the central portion of which was employed as an approximate plane wave. In this configuration, the longer focal length yields a Gaussian beam with a larger width, allowing the central region with uniform intensity distribution to serve as a plane wave.
It should be noted that the quality of the resulting intensity patterns is higher in part (a). This is attributed to the superior beam quality of the Gaussian beam produced by the doublet lens, compared to the approximated plane wave generated by the singlet lens in part (b).

\section{Conclusion}
In this work, we have proposed and demonstrated a high-efficiency method for generating PRG-based RCBs/RCBLPs under plane/Gaussian beams using a simple, low-cost, and PAH. The hologram is designed by embedding the transmission function of a binary PRG into a sinusoidal amplitude grating. This approach completely bypasses the need for complex and expensive phase-only SLMs.

Our theoretical analysis and experimental results confirm that the generated PRG-based RCBs/RCBLPs exhibit significantly enhanced power efficiency, delivering approximately five times the useful optical power in the desired orders compared to SLM-generated counterparts.
This improvement is primarily due to the elimination of diffraction-order crosstalk and power losses inherent in SLM-based methods.
In the proposed design, a phase-amplitude enhancement mechanism occurs inherently across diffraction orders: the phase amplitude of the embedded PRG is multiplied by the diffraction order number.
As a result, different, well-defined RCBs/RCBLPs are generated spontaneously in each non-zero diffraction order, while the zero order retains a plane/Gaussian profile---a feature unattainable with direct Gaussian illumination of an amplitude radial grating.

We further quantified the propagation stability of the generated RCBLPs, showing that their invariance distance is directly governed by the width of the incident Gaussian beam. A wider beam preserves the RCB-like profile over a longer range, a property we characterized using a similarity metric.

In summary, this PAH-based technique offers a superior, power-efficient, and accessible platform for generating complex structured beams. Its advantages in cost, simplicity, and performance make it particularly promising for practical applications requiring robust and efficient beam shaping, such as multi-particle optical trapping and free-space optical communication systems.

\section*{Funding.} 
Institute for Advanced Studies in Basic Sciences (G2025IASBS12632).


\section*{Disclosures.} The authors declare no conflicts of interest.

\section*{Data availability.} 
The MATLAB codes that support the findings of this article are openly available in Ref. \cite{rasouli_open-source_2026} at Figshare, with the DOI: 10.6084/m9.figshare.31398084.

\bibliography{sample}

\end{document}